\documentclass[11pt]{revtex4-2}

\usepackage{graphicx}
\usepackage{dcolumn}
\usepackage{bm}
\usepackage[table]{xcolor}
\usepackage{graphicx}
\usepackage{adjustbox}
\usepackage{placeins}
\usepackage[T1]{fontenc}
\usepackage{lipsum}
\usepackage{csquotes}
\usepackage{colortbl}
\usepackage{hyperref}
\usepackage{bm}
\usepackage{multirow}
\usepackage{booktabs}
\usepackage{epsfig}
\usepackage{amsmath,amssymb}

\begin{document}

\title{Reinforcement learning for traversing chemical structure space: Optimizing transition states and minimum energy paths of molecules}
\author{Rhyan Barrett}
\affiliation{Institute of Chemistry, Faculty of Chemistry and Mineralogy, University of Leipzig, Johannisallee 29, 04103 Leipzig}
\author{Julia Westermayr}
\email{julia.westermayr@uni-leipzig.de}
\affiliation{Institute of Chemistry, Faculty of Chemistry and Mineralogy, University of Leipzig, Johannisallee 29, 04103 Leipzig}
\affiliation{Center for Scalable Data Analytics and Artificial Intelligence (ScaDS.AI), Dresden/Leipzig, Germany}


\date{\today}


\begin{abstract}
In recent years, deep learning has made remarkable strides, surpassing human capabilities in tasks like strategy games, and it has found applications in complex domains, including protein folding. In the realm of quantum chemistry, machine learning methods have primarily served as predictive tools or design aids using generative models, while reinforcement learning remains in its early stages of exploration. This work introduces an actor-critic reinforcement learning framework suitable for diverse optimization tasks, such as searching for molecular structures with specific properties within conformational spaces. As an example, we show an implementation of this scheme for calculating minimum energy pathways of a Claisen rearrangement reaction and a number of $S_{N}2 $ reactions. Our results show that the algorithm is able to accurately predict minimum energy pathways and thus, transition states, therefore providing the first steps in using actor-critic methods to study chemical reactions.
\end{abstract}
\maketitle

\newpage



In recent years, machine learning (ML) has had a large impact on society in many different areas. From large language models \cite{kasneci2023chatgpt,kaddour2023challenges, chang2023survey} such as ChatGPT\cite{OpenAI.} to various applications in the financial sector. \citep{fang2021ascertaining,huang2020deep, wong1998neural} However, ML has only recently found its way to subject areas outside of a computational or financial setting. In physics and chemistry, ML is still in its infancy. Initial works in ML for chemistry mainly look at predicting primary outputs such as wave functions \citep{unke2021se, schutt2019unifying} or electron density \citep{jorgensen2022equivariant, grisafi2018transferable}, secondary outputs such as energies, forces or dipole moments \citep{schutt2018schnet, schutt2021equivariant, gastegger2021machine, batzner20223, smith2017ani, unke2019physnet,westermayr2020JCP} and tertiary outputs such as reaction rates \citep{komp2022progress, liu2022machine} or fundamental gaps \citep{westermayr2023high, meftahi2020machine,westermayr2021CS} of quantum systems using supervised ML. These developments have allowed scientists to calculate electronic and other properties at a much larger speed and hence lower computational cost than the associated quantum mechanical reference methods.

In a general setting, the task of predicting properties of a quantum system, or, in principle, any predictive task, involves constructing a function that maps a parameter space to a one-dimensional property space, such as energy or forces, or to a higher-dimensional property space, for example, the parameters of a wave function or excited states.\cite{Manzhos_2020,westermayr2020machine,westermayr2020combining} This function aims to closely approximate the true properties of the system. Current quantum chemical methods operate on the basis of a similar principle. In these methods, the objective is to build a parameterized wave function or probability density that minimizes the energy of the quantum system using the variational principle \cite{wang2019variational}. For each new molecular structure, another optimization of the wave function must be performed. A general function that adequately describes the mapping between the set of molecular structures and their associated properties still remains unknown.

Neural networks are known to be universal function approximators \cite{hornik1989multilayer} due to their high parameter redundancy, which allows them to construct mappings between almost any two sets. Thus, in theory, they should be able to construct this function form given the relevant dataset. Current work in quantum chemistry looks at encoding useful physical properties into neural network architectures \citep{batzner20223,thomas2018tensor,frank2022so3krates,unke2021spookynet,fuchs2020se} to produce a functional form more similar to the true mapping of molecular structures to the associated properties, thus reducing the number of parameters needing to be optimized to obtain similar errors.

The problem of finding these mappings can also be formulated slightly differently, rather than optimizing parameters with a currently existing optimizer we could look to have another neural network to find the best fitting parameters for a specific objective. To do this, one can think about the process of changing parameters as a Markov chain in parameter space, where at each step the parameters from the previous step are adapted. A neural network can then learn to explore the parameter space to find the parameters that maximize or minimize a certain property. This is the idea behind reinforcement learning. The advantage of using reinforcement learning in comparison to a normal optimization process is that a larger amount of exploration is obtained in the parameter space.
One area where the optimization of parameters is of interest is the space of different molecular structures where one wishes to optimize the positions of atoms against some reward function. Some examples are geometry optimization, minimum energy pathway calculation, or the search for critical points on excited-state potential energy surfaces, such as conical intersections. In this work, we explore the application of reinforcement learning to this task and test it on the search for minimum energy pathways and transition states.

\begin{figure}[hbt!]
	    \centering
	    \includegraphics[scale=0.5]{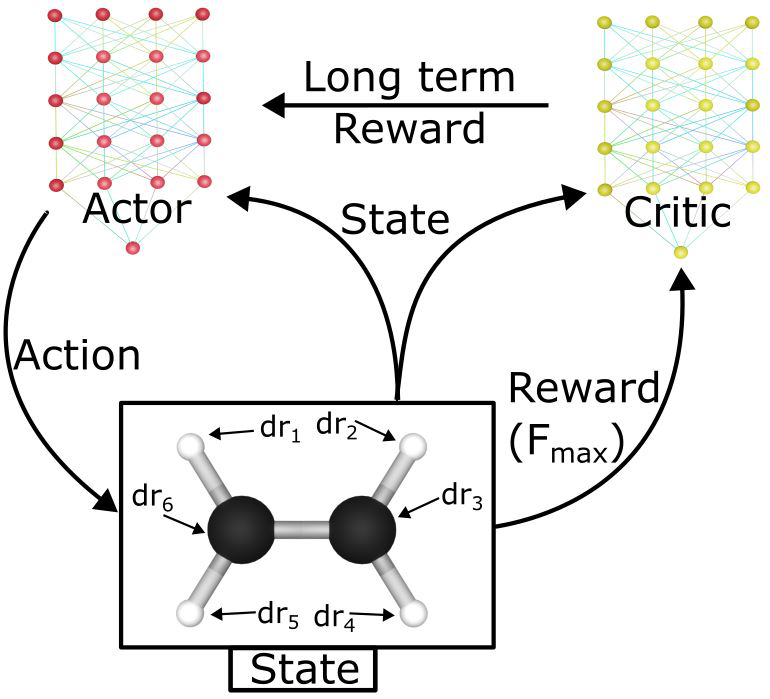}
	    \caption{\textbf{Actor-critic workflow.} The actor constructs an action which alters the positions of the atoms in the molecule. The current configuration of the molecules is known as the state. A reward is calculated using the new configuration which is passed to the critic so that the long term reward can be estimated. The long term reward and current state are then used in the optimization of the actor's policy. Given the updated policy the process is then repeated.}
        \label{fig:master}
\end{figure}

A scheme of the algorithm developed in this work is illustrated in Figure \ref{fig:master}. The exploration of a reinforcement learning algorithm can be expressed as a Markov chain. Initially, a configuration is taken and the positions are adapted with an action constructed by the neural network. The reward $r_i$ is then calculated and fed back to the neural network so that it can re-evaluate its decision to improve in the future. More precisely, the definitions of the different components that are to be defined for this task and in the context of molecular systems are summarized and described below:

\begin{itemize}
\item \textbf{State}: Current conformation of an $m$-atomic molecule described by atomic numbers, $Z_i$, and positions, $R_i \in \mathbf{R}^3$:
\begin{equation}
   S_t = \{(Z_0, R_0), (Z_1, R_1), ...., (Z_m,R_m) \}.
\end{equation}

\item \textbf{Action}: Given the atomic numbers and positions in the state, the set of corresponding new positions can be constructed. The new positions, $R_{i} + \delta R_{i}$, are then constructed as follows:
\begin{equation}
\{ R_{0} + \delta R_{0}, R_{1} + \delta R_{1},.., R_{i} + \delta R_{i},.., R_{m} + \delta R_{m} \}  \:\:\:\: i \in (0,1,,,,m).
\end{equation}
To maintain symmetry constraints, a switch to internal coordinates may be conducted.

\item \textbf{Reward}: The reward is used to let the reinforcement learning agent assess the success of its actions and is dependent on the particular task at hand. In the case of geometry optimizations, the task would be for instance to maximize $E_t^{-1}$ to obtain the minimum energy structure. 
\end{itemize}
The reward calculated is defined as the immediate feedback after a change in the current state. However, in this case, the quantity of interest is the long term reward or the reward given by the final structure. As the long term reward is not known until the task is complete, an estimation of an expected long term reward given by the current state is calculated. Two main methodologies in reinforcement learning have been developed for this purpose \cite{arulkumaran2017deep}, namely value-based learning, such as Q-learning \cite{watkins1992q}, and policy based methods, such as actor-critic methods \cite{konda1999actor}.  

The Actor-Critic method \citep{grondman2012survey,mnih2016asynchronous} used in this work is a class of reinforcement learning shown to be effective in high dimensional problems \citep{zhou2021actor,zhang2020actor}, with which will be dealt with in many situations in chemistry.
This method involves three components: an environmental state, an actor that interacts with the environment, and a critic. This concept is illustrated in Figure \ref{fig:master}. The objective of the actor is to develop a policy, $\pi(s|\theta)$, to maximize long term rewards and the critic attempts to predict the expected long term reward, $V(S_{k})$. 

More precisely, looking at each step the actor receives a state that is the result of the action performed on the environment. In this work, the action consists of changes to the current positions of a molecule, the state. The reward of the given state is calculated using some evaluator, \textit{i.e.}, the energy of the new state or the corresponding forces. This reward along with the state is passed to the critic, which then estimates what the expected long term reward will be given the current state.  
The expected long term reward is the core quantity in actor-critic models from which the actor learns to adjust itself. However, this quantity is not straightforward to calculate as it would be needed to calculate all possible moves from a particular state to then take the final reward.To avoid this tedious task, a neural network is used to estimate this quantity.
\begin{figure}[hbt!]
	    \centering
	    \includegraphics[width=0.5\columnwidth]{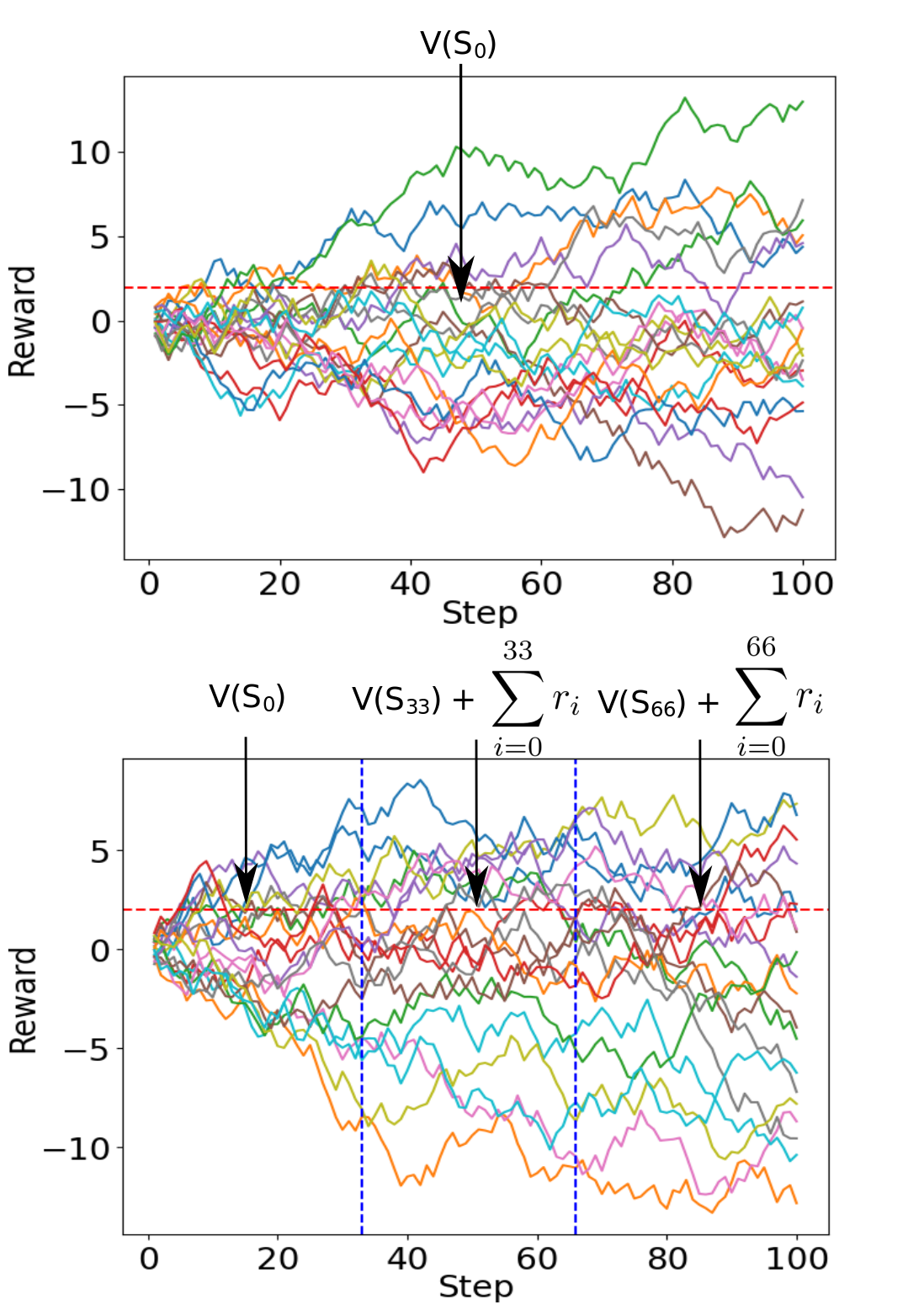}
	    \caption{\textbf{Possible rewards generated by an actor-critic algorithm.} a) The solid lines represent how the reward evolves over the episode and the dotted line represents the critic's estimate at step $S_0$, $V(S_0)$.  b) Again the solid lines represent how the reward evolves over the whole episode. The episode can then be broken up into several pieces, in each segment an estimation of the expectation is calculated. In the first segment the expectation is the critic's estimate at state $S_0$, at step $S_{33}$ it is given by the critic estimate at step $33$ plus the previous rewards and finally in the final segment the estimate is the critics estimate at step $66$ plus the previous rewards.}
        \label{fig:reward}
\end{figure}

Figure \ref{fig:reward} illustrates an example of the rewards generated by a set of random walks. The random walks show how an actor might change the position of the single atom along with the associated rewards from an initial state $S_k$. The straight line represents the critics estimate of the rewards given the previous actions the actor has performed; if the starting state is $S_k$ then this value is written as $V(S_k)$. However, this scheme can prove quite inefficient since every critic update must be performed at the end of the episode. To provide an approximation of $V(S_k)$ before the end of the episode, temporal difference learning is used.
 Temporal difference learning is illustrated in Figure 3b. It involves breaking down an episode into smaller pieces of size $m$, then looking at the final reward received plus the estimate from the critic at the final state. In this way, the actor and critic are able to update themselves more frequently. 
 
To define mathematically, the cumulative rewards received through the episode, $R_k$,  can be rewritten:
\begin{equation}
     R_k = \sum_{t=k}^{n} r_t  =  \sum_{t=k}^{k+m} r_t + \sum_{t=k+m}^{n}  r_t =  \sum_{t=k}^{k+m} r_t + V(S_{k+m}),
 \end{equation}
with $r_t$ being the rewards of each individual step.  As can be seen above the total rewards received are split through the episode into two sums, one for the smaller episode piece of length $m$ and then the sum for the remainder of the episode which is approximated by the critic. This new estimate is used to update the critic. Due to the output of the critic being used to adjust the critic in the training process, actor-critic methods can become unstable so correctly selecting hyper-parameters is critical for proper training and preventing divergence in the loss.

Now that an estimate of the long term rewards of a particular molecular state has been obtained it can be integrated with the agent or actor so that better decisions are obtained. As mentioned earlier there are two main methods of doing this but the focus will be on policy based methods namely actor-critic methods. Actor-critic methods operate by taking the estimated long term reward using the critic and measuring if the decisions of the actor lead to a value higher or lower than the expected long term reward. The advantage can be written as

\begin{equation}
    A_{k+n} = R_k - V(S_k).
\end{equation}

If $A_{k+n}>0$ then the series of actions taken by the actor can be considered as positive since they lead to rewards that were above average. This in turn leads to these actions being positively reinforced in the actor. On the contrary if $A_{k+n}<0$  this corresponds to the actor selecting actions which lead to the rewards being below average. This changes the actors behaviour in order to reduce the probability that these actions happen again. Over time the actor should learn the steps which lead the maximization of the reward function by selecting actions with positive advantages.

Now that the overall structure of the reinforcement learning algorithm has been described, we can consider how to implement this in the case of molecular structures. The decisions of the actor can be either deterministic or stochastic in nature, \textit{i.e.}, either the actor output is the new positions of the atoms or a probability distribution from which a new position is sampled. The second case is preferable since it enforces more exploration and thus a higher likelihood of finding an optimal solution. Let us introduce this mathematically, define $R^{t}_{i}$ as the position of atom $i$ in the molecule at step $t$ in the episode then the position of atom $i$ at time step $t+1$ is sampled from the distribution $P(R_i^t | R_i^{t+1}, \pi(s|\theta))$. The probability distribution for the next configuration in the episode at time step $t$ can then be written as,
\begin{equation}
  P(R_t)  =  {\alpha} \prod_{i=0}^{N_{atoms}} P(R_i^t | R_i^{t+1}, \pi(s|\theta))
\end{equation}
Here, $\alpha$ is the normalisation constant. The positions can then be sampled from the generated probability distributions to construct the new conformer. This process is repeated to form a Markov chain. 

To reiterate, the critic must estimate the long term rewards $V(S_k)$ but also be permutation invariant of the input by considering each atom individually then summing their associated contributions. Thus $V(S_k)$ can expanded as follows,
\begin{equation}
V(S_k) = \mathop{\mathbb{E}}(R_k| S_k = S) = \mathop{\mathbb{E}} \left[ \sum_{t=k}^{n} r_t | S_k = S \right] = \sum_{i=0}^{N_{atoms}} \mathop{\mathbb{E}} \left[ \sum_{t=k}^{n} r_t^i | S_t = S \right].
\end{equation}
where summing over $r_t^i$ corresponds to each atom's estimated contribution to the total reward at time step $t$. This structure allows easy implementation of the critic as a neural network for arbitrary-sized molecules. Now the overview of an actor and critic system is given we will attempt to apply this to the case of minimum energy pathway prediction as an application of actor-critic methods in a molecular setting.

Conventional methods to compute minimum energy pathways require many sequential evaluations of the quantum Hamiltonian leading to high computational costs and due to the vastness and intricate local topological structure of the potential energy surface, usually requiring lots of human input to successfully converge. Therefore, ML is promising to advance this field, but only few works exist in this direction\citep{jackson2021tsnet, zhang2021deep,choi2023prediction,duan2023accurate}. One work is TS-Net \cite{jackson2021tsnet}, which applies a tensor-field network to predict the structure of a transition state. However, this method requires a training set with transition states and the generation of transition states for a training set is computationally expensive and time-consuming, thus limiting the applicability of this model, especially when targeting large systems. Another work reformulates the transition state search into a shooting game using reinforcement learning techniques \cite{zhang2021deep}. This technique is related to a common computational workflow, namely, transition path sampling, and operates by choosing a coordinate in phase-space from which two trajectories are started with opposite momenta. If trajectories reach the desired products and reactants, the episode is considered successful. While this approach is powerful in theory, it requires Monte-Carlo techniques \cite{shapiro2003monte} to identify promising pathways before training. As a consequence this method can become highly expensive as the molecular systems become larger. Thus, to study larger systems, a faster way of performing quantum mechanical calculations is required along with a way to intelligently search the potential energy surface.

To improve on the methods mentioned, the developed model is based on the frequently used nudged elastic band method (NEB) \citep{jonsson1998nudged, henkelman2000climbing} to predict minimum energy pathways. NEB involves dividing the reaction pathway into a series of discrete images or "beads". Each bead represents a possible intermediate state. Given a reaction pathway, the force on each image is determined by a combination of the internal inter-atomic forces of each image perpendicular to the reaction pathway, $F_{int}\perp$, and the virtual harmonic spring forces holding the images together parallel to the reaction pathway, $F_{spr} \parallel$.  More precisely, the set of forces on the atoms in image $j$ is given by:
\begin{equation}
F^{i} = F^{i}_{spr} \parallel + F_{int}\perp =  F^{i}_{spr} \parallel - \nabla E(R_i) \perp  \end{equation}
\begin{equation}
F^{i}_{spr} \parallel = k(|R_{i+1} - R_{i}| - |R_{i} - R_{i-1}|)\tau_i F^{i} = F^{i}_{spr} \parallel + F_{int}\perp =  F^{i}_{spr} \parallel - \nabla E(R_i) \perp  \end{equation}
\begin{equation}
\nabla E(R_i) \perp = \nabla E(R_i) - \nabla E(R_i) \parallel
\end{equation}

Here $R_i$ represents the positions of atom $i$ in image $j$, $\tau_i$ is the tangent to the reaction path at image $j$ and $\textit{k}$ is the spring constant controlling the strength of the harmonic springs. Using this, the maximum force $F_{max}$ on any one atom in any image along the reaction pathway can be computed. Here the aim is to optimize the molecular pathways by adapting molecular configurations in a way that minimizes $F_{max}$.

Keeping in mind the NEB algorithm and the actor-critic model derived above, the task of finding the minimum energy pathway can be formulated as a Markov process with each state being formed by a series of molecular conformations along the reaction path. In a similar way to the NEB method, at each step in the process the positions of the molecules on the reaction pathway are modified. This differs from the previous derivations since a set of molecular structures needs to be considered in contrast to a single molecular structure.
Therefore, the previously defined formula need to be rewritten such that they account for multiple geometries. The full derivation of this will be left in the supplementary material, see section S5. Due to the extremely large configuration space made up by the images along the reaction pathway, a subset of configurations is constructed, which provides a good starting point. To do this the $F_{int}\perp$ and $F_{spr} \parallel$ values for the atoms in each image are constructed using a PaiNN model \cite{schutt2021equivariant}. Then for each image a linear combination of these values is generated to form the desired subspace of potential moves. Analogous to above, the reinforcement learning algorithm can again be broken down into a series of components. The state is represented by molecules along the reaction path described by atomic numbers and positions with each reaction path being described by $n$-interpolated images. Thus, the state space consists of $n$ molecules with their associated positions and atomic numbers. The action as mentioned before consists of a linear interpolation of the vectors $F_{int}\perp$ and $F_{spr} \parallel$ for an individual image. The new positions can be constructed from this. The reward can be defined as the maximum force $F_{max}$, as shown above. The reward, $r_t$, at each step is given by $(F_{max})^{-1}$. Maximizing the given reward and minimizing the $F_{max}$ value are equivalent.

In order to achieve an accurate description of the policy and value functions, it is important to convert the Cartesian coordinates into a representation that is both rotationally and translationally invariant \cite{schutt2021equivariant}. This can be accomplished through the use of the PaiNN representation developed by Sch\"{u}tt \textit{et al.}\cite{schutt2021equivariant}. Based on this representation, the aim is to predict changes to the positions of atoms in each image along the reaction pathway at each step with the goal to move closer to the minimum energy pathway. Around each atom in the molecule, a fine discrete grid of $\alpha_i$ and $\beta_i$ values is constructed to produce a linear combination of $F_{int}\perp$ and $F_{spr} \parallel$. A distribution is then constructed by the model on this grid and the new position is sampled from the respective distribution. An illustration of this and how the associated model is built can be seen in Figure \ref{fig:actor}. Further details of this can be found in section S2 in the supplementary material. Additionally details of the training procedure and loss can be found in section S1 of the supplementary material.

\begin{figure}[hbt!]
\includegraphics[scale=0.2]{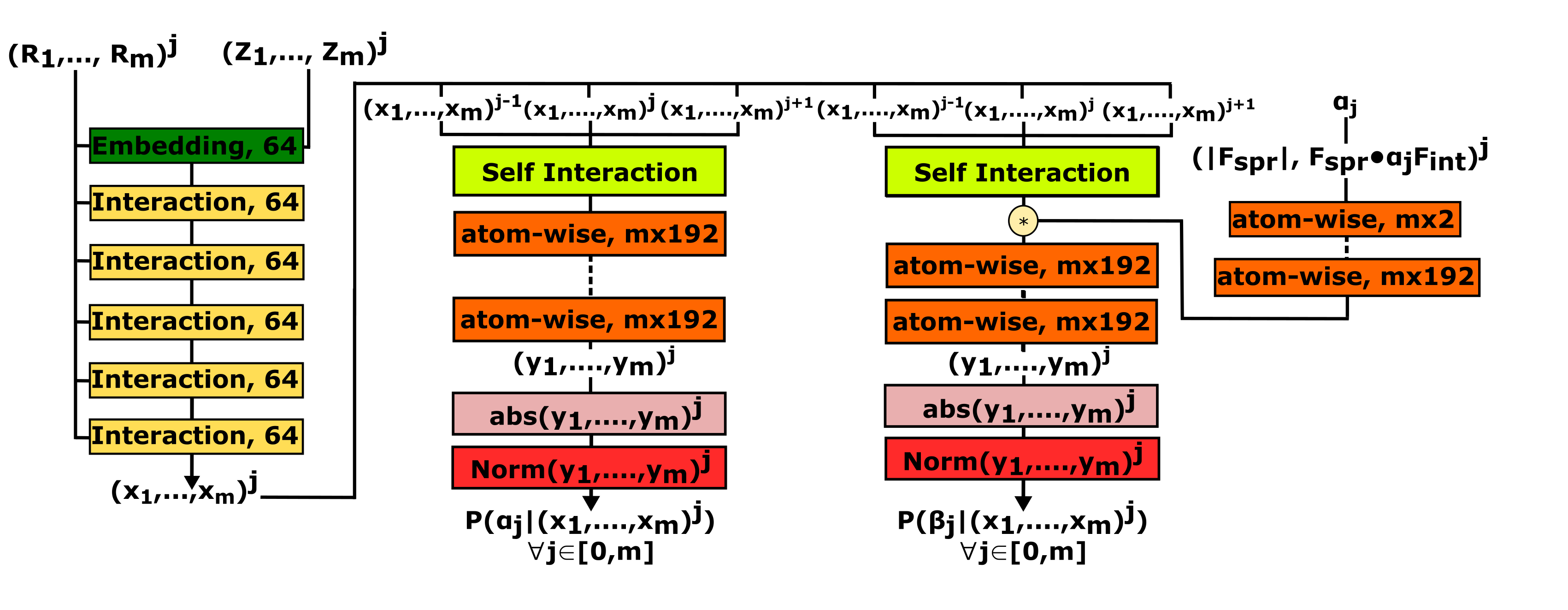}
\caption{\textbf{Actor architecture.} The PaiNN representation forms the initial feature vectors for the actor. The positions of each atom in the current image are passed through a self interaction layer and a dense layer to produce the $\alpha_i$ distribution. A similar process is used to construct the associated $\beta_i$ distribution. The two distributions are then sampled from to produce the change in positions at the current time step.}
\label{fig:actor}
\end{figure}

To incorporate symmetry considerations into the predictions, the outputs of a series of PaiNN layers are utilized. These layers form the initial feature set that is subsequently integrated into the remainder of the model. Next, a self-interaction layer \cite{vaswani2017attention} is added to condense the information between neighboring images along the reaction pathway into a single feature vector. These feature vectors are then passed through a series of atom-wise layers. A distribution for $\alpha_i$ is produced from which values are sampled. These are used to construct an additional feature vector containing information about the magnitude of $F_{int}\perp$ along with an associated dot product with $F_{spr} \parallel$. This feature vector along with the feature vector produced by the self-interaction layer are combined together using a point-wise multiplication followed by a series of atom-wise layers. The associated $\beta_i$ distribution is then produced and sampled from in order to construct the new position. An illustration of this can be seen in Figure \ref{fig:actor}.

The critic model shares the same PaiNN representation layers as the actor model but instead the final layers are used to provide an estimate of the value function in a similar way as mentioned before.
In the final layers of the critic we sum over all the contributions for each individual atom to get the estimate. An overview of the structure is shown in Figure S1 in the supplementary material including a more detailed description in section S3.


The performance of the method is tested on a series of test reactions, namely the allyl-\textit{p}-tolyl ether Claisen rearrangement reaction and on multiple reactions obtained from the $S_{N}2$ data set comprising chemical reactions of the form $X^{\text{–}} + H_3C\text{–}Y \rightarrow X\text{–}CH_3 + Y^\text{–}$ with $X, Y \in \{F, Cl, Br, I\}$ and $X\neq Y$.

\begin{figure}[hbt!]
\includegraphics[scale=0.2]{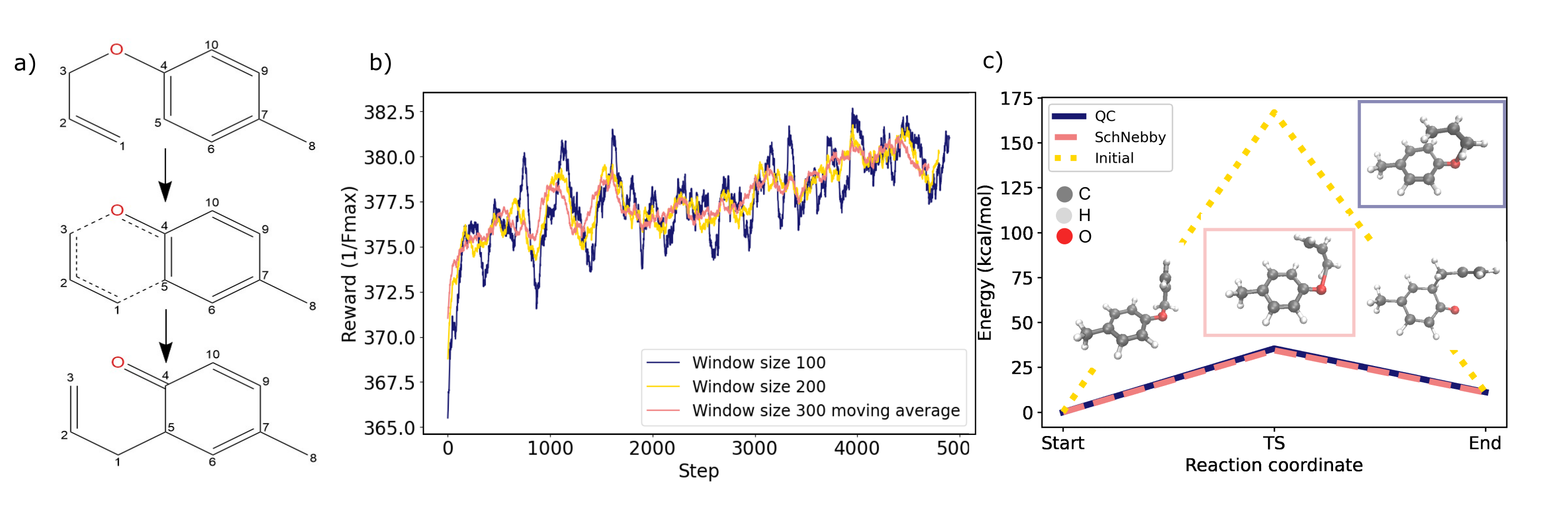}
\caption{\textbf{Claisen rearrangement reaction.} a) An illustration of the start, intermediate and end structures in the allyl-\textit{p}-tolyl ether Claisen rearrangement reaction. b) Rewards and episode number over the course of the training phase representing the moving average rewards values with 3 respective  window sizes. c) Energies and structures of start and end geometry and found transition state (TS) obtained with quantum chemistry (QC, blue), actor-critic model (red), and the initial guess (yellow) from geodesic interpolation.}
\label{fig:claisen}
\end{figure}

First, the model is tasked to target the pathway of the allyl-\textit{p}-tolyl ether Claisen rearrangement reaction. As illustrated in Fig. \ref{fig:claisen}a, the Claisen rearrangement takes place through a concerted mechanism in which a C-C bond forms between the C1 of the allyl group and the ortho position of the benzene ring (marked as C5) at the same time that the C3-O bond of the ether breaks. This rearrangement initially produces a non-aromatic intermediate which quickly undergoes a proton shift to reform the aromatic ring in the product. Claisen rearrangement occurs in a six-membered, cyclic transition state involving the concerted movement of six bonding electrons in the first step.  The full path of the reaction is attached as supporting information. 

To assess the reward, a PaiNN\cite{schutt2021equivariant} model was trained on a dataset containing structures of allyl-\textit{p}-tolyl ether obtained from metadynamics simulations taken from ref.\citenum{gastegger2021machine} to form the initial representation input for the model. The mean absolute errors (MAEs) for energies and forces are 0.22 kcal/mol and 
0.37 kcal/mol/\AA  ~, respectively, on a hold-out test set (details on training of PaiNN models are in the supplementary material in section S4.1-S4.2 and Figure S2). Following from this, an initial guess is used of the reaction pathway obtained via geodesic interpolation. For consideration of computational efficiency, the model is allowed to sample for $10$ episodes of length $50$ to find a pathway of lower $F_{max}$ value as the new initial starting guess. The agent is now trained from this starting guess to minimize the $F_{max}$ values. Exact implementation details can be found in the supplementary material in section S4.4.

The training process can be followed in Figure \ref{fig:claisen}b that illustrates a consistent increase in the reward function and thus a decrease in the associated $F_{max}$ value. Figure $\ref{fig:claisen}$c shows the different energy curves and transition states found with the model and quantum chemistry (QC) using standard NEB with DFT at PBE0-D4/def2-TZVP level of theory. The activation energy obtained using the model was calculated to be about $37$ kcal/mol, which is in very good agreement to the reference value of $35$ kcal/mol, especially when considering the initial guess of over 150 kcal/mol. 

In the second task, the aim is to minimize the $F_{max}$ value of a series of $S_{N}2$ reactions and predict the associated transition state structures. The $S_{N}2$ reactions under consideration are as mentioned, reactions of the following form:
$X^{\text{–}} + H_3C\text{–}Y \rightarrow X\text{–}CH_3 + Y^\text{–}$ with $X, Y \in \{F, Cl, Br, I\}$ and $X\neq Y$. Again, the reward is computed using PaiNN. The MAEs for energies and forces are 0.87 kcal/mol and
0.20 kcal/mol/\AA  ~,respectively, on a hold-out test set (see section S4.3 and Figure S3 for further details).

In contrast to the previous experiment, no initial sampling is done by the model prior to the training period, thus the episodes commence initially from the geodesic interpolation. Again for computational efficiency, in each episode uses $10$ steps. 
\begin{figure}[hbt!]
	    \centering
	    \includegraphics[width=0.7\columnwidth]{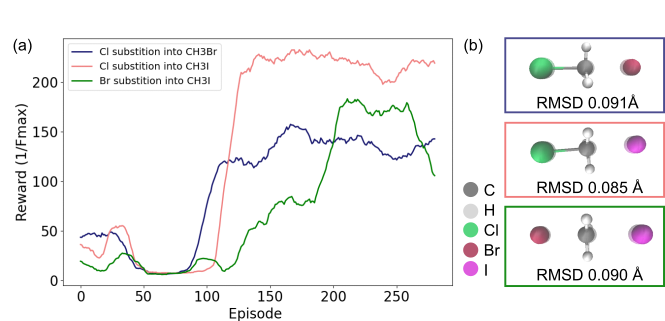}
	    \caption{\textbf{$S_N2$ reactions.} a) Rewards through a training period of $300$ episodes for a series of $S_{N}2$ substitution reactions. b) Transition states obtained with the model (solid) compared to transition states obtained with the original quantum chemical method (transparent) including root mean squared deviations (RMSD). }
        \label{fig:sn2}
\end{figure}
Looking at Figure\ref{fig:sn2}a, in all cases the reward is maximized and the associated $F_{max}$ is minimized. In order to give a comparative view on the effectiveness of the model, the transition structures produced by the model are compared with the reference structures. The structures are plotted on top of each other in Figure \ref{fig:sn2}b. Transparent are the structures obtained with quantum chemistry. As can be seen, the structures are in excellent agreement to each other. For quantitative measure, further computation of the root mean squared deviations (RMSDs) is performed, which are shown below the images and are below 0.1 \AA in all cases.

The test cases show the use of the model to predict energy pathways and transition states of a series of reaction mechanisms as demonstrated for the organic allyl-\textit{p}-tolyl ether Claisen rearrangement reaction and $S_{N}2$ substitution reactions. The results show that the model produces transition states and corresponding energy curves, which closely resemble the quantum chemical reference values. While the model can efficiently be trained on single reaction, future research is needed to assess the performance on training of many diverse reactions at once. Further effort will thus be devoted to the development of expanding the model to generalize more easily to a whole set of reactions. Additionally further research will also look into integrating actor-critic methods into other molecular tasks.
In addition, the restriction to hyper planes generated by the internal and spring forces can be removed to allow for larger search space but with a larger computational cost. As the reward function can be changed depending on the use case to allow the actor-critic algorithm to target conformations with a certain set of properties, we expect that the reinforcement learning model has the potential to become a valuable tool not only for the estimation of transition states and minimum energy paths, but also can advance the search for molecular conformations with target properties.

\section{Code Availability}
The code is publicly available at: https://github.com/rhyan10/\_SchNebby\_.
\section*{Data availability}

The dataset used for the Claisen rearrangement reaction is publicly available at \citenum{claisendata} under the name ate\_vacuum.tgz. 
The dataset for the $S_N 2$ reactions can is publicly available at \citenum{sn2data}.

\section{Acknowledgements}
This work is funded in parts by the Deutsche Forschungsgemeinschaft (DFG) -- Project-ID 443871192 - GRK 2721: "Hydrogen Isotopes $^{1,2,3}$H". The authors acknowledge the ZIH TU Dresden and the URZ Leipzig University for providing the computational resources. We thank Jakob Schramm for quantum chemical reference calculations (NEB) of the allyl-\textit{p}-tolyl ether Claisen rearrangement reaction and Dr. Michael Gastegger, Dr. Oliver Unke, and Prof. Ralf Tonner-Zech for fruitful discussions regarding the project. Additionally we thank Hendrik Weiske, Toni Oestereich and Luisa Kärmer for their help.

\section*{Author Contributions}
R.B. and J.W. planned and designed the project. R.B. implemented the algorithms and models. R.B. tested the models. R.B. and J.W. were involved in various discussions throughout the project, wrote and refined the manuscript.

\section*{Conflicts of interest}
There are no competing interests to declare.

%

\end{document}


\title{Supplementary material for: Reinforcement learning for traversing chemical structure space: Optimizing transition states and minimum energy paths of molecules}
\author{Rhyan Barrett}
\affiliation{Institute of Chemistry, Faculty of Chemistry and Mineralogy, University of Leipzig, Johannisallee 29, 04103 Leipzig}
\author{Julia Westermayr}
\email{julia.westermayr@uni-leipzig.de}
\affiliation{Institute of Chemistry, Faculty of Chemistry and Mineralogy, University of Leipzig, Johannisallee 29, 04103 Leipzig}
\affiliation{Center for Scalable Data Analytics and Artificial Intelligence (ScaDS.AI), Dresden/Leipzig, Germany}

\date{\today}
\maketitle
\tableofcontents
\newpage
\section{Loss Implementation}

To reiterate from the paper $A_{k+n}>0$ and $A_{k+n}<0$ correspond to the associated action being positively and negatively reinforced, respectively. The weight updates of a general actor are then constructed by weighting the policy gradients of each sample by their advantage value,
\begin{equation}
\nabla_{\theta} \mathcal{L}_{a} = \nabla_\theta V(S_{k}) = \mathop{\mathbb{E}}_{\pi_\theta} [\nabla_\theta \pi_\theta(s|a) A_{\pi_\theta} (s,a) ]
\end{equation}
The loss, $\mathcal{L}_c$ and weight updates for a general critic are calculated using a mean squared error (MSE) between the observed reward function and predicted value function $V(S_k)$:
\begin{equation}
\mathcal{L}_{c} = \left( \sum_{t=k}^{k+n} \gamma^{t} r_t + V(S_{k+n}) - V(S_{k})\right).
\label{eq:Lc1}
\end{equation}

The training process in our actor-critic implementation is described as follows. The loss is computed using the advantage values obtained from the temporal difference targets (TD-targets). The total loss is equal to the sum of the actor and critic loss, $\mathcal{L}_a$ and $\mathcal{L}_c$, respectively:
\begin{equation}
\mathcal{L}_{total}=\mathcal{L}_a + \mathcal{L}_c
    \label{eq:losstotal}
\end{equation}
The total loss function for the actor, $\mathcal{L}_{policy}$ is calculated using an adapted version of the loss as mentioned below. The value $\mathcal{L}_{policy}$ weights the policy gradients calculated by the advantages of those actions. The entropy loss, \textit{i.e.}, the second part in equation (4), encourages exploration by penalizing overly confident predictions. To this end we utilize a hyper-parameter $\beta$ controlling the strength of the penalty of the entropy loss in order to control the amount of exploration. 
\begin{equation}
\begin{split}
\mathcal{L}_{policy}(action_{Pos}) \\ = \sum_{k=1}^{samples} \left( \sum_{i=1}^{n} A_{k,i} \: log(Softmax(t_{k,i})) + \beta \sum_{i=1}^{n} t_{k,i}*log(Softmax(t_{k,i})) \right)
\label{eq:Lpolicy}
\end{split}
\end{equation}
Here $t_{i,k}$ and $A_{k,i}$ corresponds to probability associated to the action at element $k$ in the batch and at step $i$ in the episode.
Since the actions correspond to position changes $\alpha_i$ and $\beta_i$, values which are similar to other $\alpha_i$ and $\beta_i$ pairs will likely correspond to similar rewards. Thus for each action chosen we weight the corresponding policy gradients by their distance to the chosen action. More precisely the final actor loss policy for a single sample becomes:
\begin{equation}
\begin{split}
\sum_{i=1}^{n} A_{k,i} \sum_{a=1}^{N_a} \frac{1}{1+|a'_{k,i}-a_{k,i}|}log(Softmax(t_{k,i,a})) \: + \\  \beta \sum_{i=1}^{n} t_{k,i}*log(Softmax(t_{k,i})) 
\label{eq:Lpolicy2}
\end{split}
\end{equation}
The values $a'_{k,i}$ and $a_{k,i}$ are the chosen action and an arbitrary action respectively. $N_a$ corresponds to the total number of actions. Additionally, to generate an initial starting guess for the reaction pathway we use geodesic interpolation followed by high speed sampling using the model's initial distribution over the actions. This produces a guess much closer to the minimum energy pathway meaning less training time is wasted on pathways which are not useful.
The critic is trained using standard temporal difference learning. In this case, the rewards are obtained from the $(F_{max})^{-1}$ values.

\section{Actor Implementation Details}

Based on the representation mentioned in the paper, we aim to predict changes to the positions of atoms in each image along the reaction pathway at each step in order to move closer to the minimum energy pathway. To achieve changes in atomic positions, we look to generate distributions for positions changes around each atom. To elaborate further, around each atom in the molecule, we construct a fine discrete grid of $\alpha_i$ and $\beta_i$ values to produce a linear combination of $F_{int}\perp$ and $F_{spr} \parallel$. A distribution is then constructed by the model on this grid and the new position is sampled from the respective distribution. It is important to note these $\alpha_i$ and $\beta_i$ values can be generated per atom in each image or be the same for every atom in the image with the former being the most computationally intense. In the following section we will describe the case where an $\alpha_i$ and $\beta_i$ value are generated for each individual atom.

In more detail, let us take the initial state $S_t$, of a reaction pathway, which contains the atomic types $\{Z_i\}$ and associated positions $\{\{R_{i_1}\}^1, \{R_{i_2}\}^2 ,..., \\ \{R_{i_n}\}^n\}$ of all atoms in each molecule along the reaction pathway where $n$ is the number of images and $i_j$ corresponds to the index of the atom in image $j$. The probability distributions of the positions of the atoms in the reaction pathway at step $T$ is given by $\{\{P(R_i)\}_j \: \forall j \in [0,n] \}$. To accommodate the auto-regressive nature of the reinforcement model we can write this as the product of conditional probabilities of the positions at each step which can be decomposed as changes in positions: 
\begin{equation}
  \{P(R_i)\}_j  =  \left[ \prod_{t=0}^{T} \prod_{i=0}^{N_{atoms}} P(R^{t-1}_{i,j} + \delta R^{t}_{i,j} | R^{t-1}_{i,j-1} , R^{t-1}_{i,j}, R^{t-1}_{i,j+1} )\right]_j \forall j \in [0,n].
\end{equation}

 From now on we will abbreviate $R^{t-1}_{i,j-1} , R^{t-1}_{i,j}, R^{t-1}_{i,j+1}$ with $Y^{t-1}_{i,j}$. Additionally, we can decompose the distribution into the joint probability distribution of $\alpha_i$ and $\beta_i$ values from which the changes in positions for each molecule along the reaction path are constructed:
\begin{equation}
\left[ \prod_{t=0}^{T} \prod_{i=0}^{N_{atoms}} P( \delta R^{t}_{i,j} | Y^{t-1}_{i,j} )\right]_j =  \left[ \prod_{t=0}^{T} \prod_{i=0}^{N_{atoms}} P(\alpha_i, \beta_i | Y^{t-1}_{i,j} )\right]_j \forall j \in [0,n] 
\end{equation}
Furthermore, we add a dependency of $\beta_i$ on the $\alpha_i$ value chosen. This leads to the joint distribution being rewritten so that $\beta_i$ depends on $\alpha_i$;
\begin{equation}
\begin{split}
 \left[ \prod_{t=0}^{T} \prod_{i=0}^{N_{atoms}} P(\alpha_i, \beta_i | Y^{t-1}_{i,j} )\right]_j  =  \left[ \prod_{t=0}^{T} \prod_{i=0}^{N_{atoms}} P(\beta_i | Y^{t-1}_{i,j}, \alpha_i ) P(\alpha_i | Y^{t-1}_{i,j})\right]_j \\ \forall j \in [0,n]. 
\end{split}
\end{equation}
These distributions are produced as an output of the actor model and combined into a categorical distribution over a fine grid. A new position is then sampled from this grid for each atom in every image. The methodology is then implemented into a deep learning architecture as seen in the paper.

\section{Critic Implementation Details}
As mentioned in the paper, the critic model employs identical PaiNN representation layers as the actor model, with the exception that the final layers are utilized to furnish an estimation of the value function
\begin{equation}
V^{\pi}(S_k) = \mathop{\mathbb{E}} _{\pi}(R_k| S_k = S) = \mathop{\mathbb{E}}_{\pi} \left[ \sum_{t=k}^{\infty} \gamma^t r_t | S_k = S \right].
\end{equation}

At each step in the auto-regressive structure of the RL algorithm, the reward is given from the NEB calculation using the $F_{max}$ quantity and a conventional PaiNN model. The atom and image of which $F_{max}$ occurs in future steps is highly uncertain. However, it can be decomposed as the probability of each atom causing the largest force multiplied by the estimated force if this is the case; the expected $F_{max}$ for a particular atom. More formally the value function decomposes as follows:

\begin{equation}
\mathop{\mathbb{E}}_{\pi} \left[ \sum_{t=K}^{\infty} \gamma^t r_t | S_t = S \right] =    \sum_{j=0}^n \sum_{i=0}^{N_{atoms}} \left[ \sum_{t=k}^{\infty} \gamma^t \mathop{\mathbb{E}}_{\pi}
(f_{max}^{z} | S_t = S) \right].
\end{equation}

\begin{figure}[hbt!]
\includegraphics[width=\textwidth]{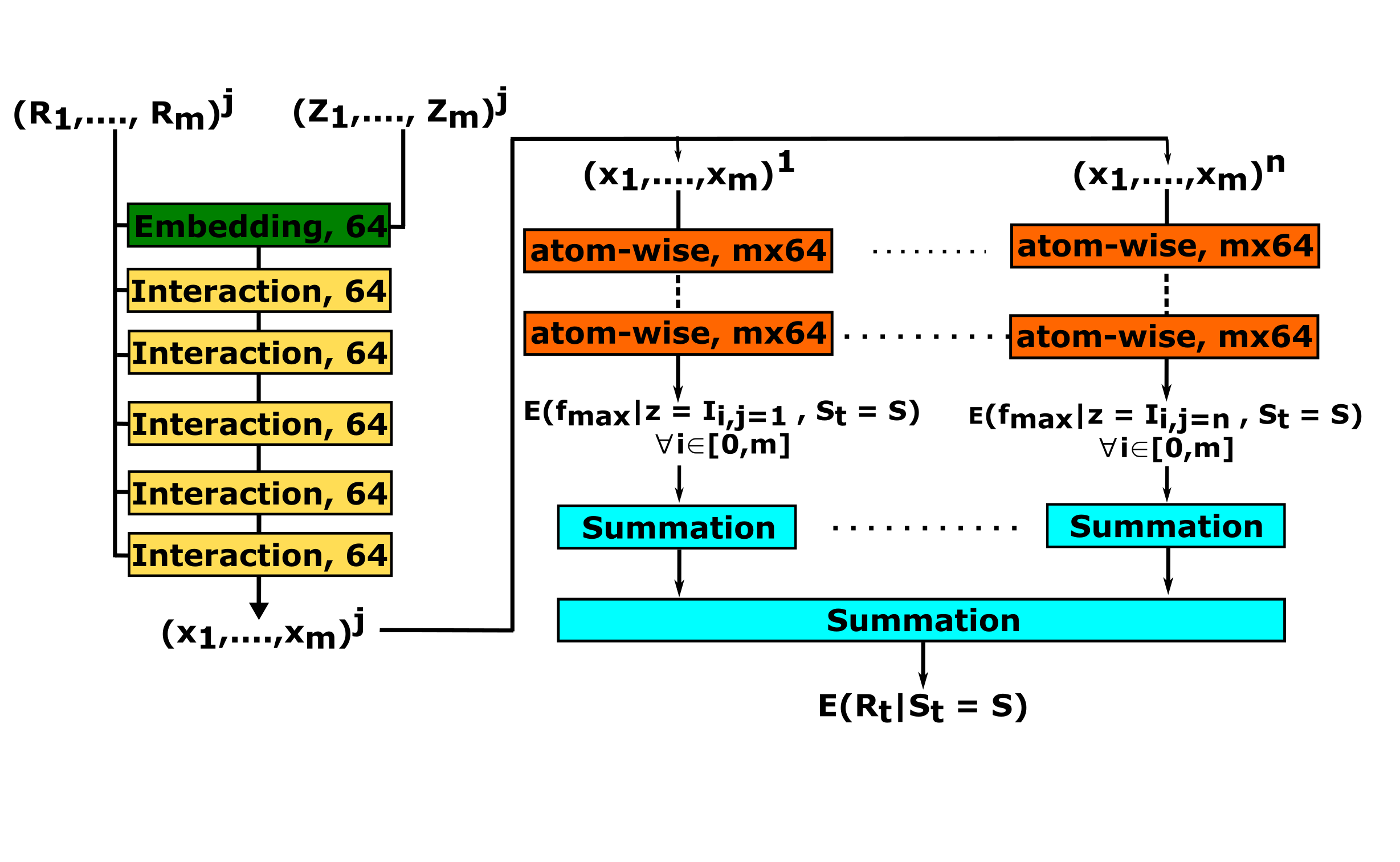}
\caption{ \textbf{Critic architecture.} The PaiNN model provides an initial feature representation, $(\mathbf{x_1},...,\mathbf{x_m})^j$ with $j$ referring to the index of the image. For each image in the reaction path we pass the feature vectors for each atom through a series of atom-wise layers. The output contains the expectation of each atom in a molecule. The expectations are then summed over all atoms in each image and finally over all images to deliver the final expectation.}
\label{fig:critic}
\end{figure}

Using the shared representation layer between the actor and critic models we then predict the expected values of each atom in each image, as it is shown in Figure \ref{fig:critic}. The resulting expectations are then summed in the final layers to give an estimate of the value function.

\section{Hyperparameters and Training}
All experiments and trainings were run on an NVIDIA RTX A2000 12 GB GPU as well as 20 i7-12700 12th Gen CPUs. 

\subsection{PaiNN Model Training Details}
 The datasets used for the training of the models are freely available and comprise the allyl-\textit{p}-tolyl ether Claisen rearrangement reaction \cite{gastegger2021machine} as well as $S_{N}2$ reactions\cite{unke2019physnet}. The data were generated with  PBE0/def2-TZVP \cite{adamo1999toward, weigend2005balanced} and DSD-BLYP-D3(BJ)/def2-TZVP \cite{kozuch2010dsd, goerigk2017comprehensive} level of theory, respectively. The target properties used for training of the Claisen rearrangement and $S_{N}2$ reactions were energy and forces. The loss weighting of energy and forces was 0.02 and 0.98, respectively, for both datasets. The dataset was split into training, validation, and test sets randomly. For the claisen rearragement reaction, 50000 data points were used for training, 6000 for validation, and 5000 for testing. We used a batch size of 50. For the $S_N 2$ dataset, 350000 data points were used for training, 50000 for validation, and 52709 for testing. A batch size of 500 was used. 

For both datasets a cutoff function was applied. In addition, for the Claisen rearrangement reaction, the means of each target property were substracted from the dataset as well.
Both models were built using SchNetpack \cite{schutt2022schnetpack} in combination with the equivariant PaiNN \cite{schutt2021equivariant} representation. The hyperparameters for the model used in both the Claisen rearrangement and $S_N 2$ reactions are as stated in Table 1.
\begin{table}[hbt!]
\begin{center}
\caption{Hyperparameters of the final PaiNN models trained on the Claisen rearrangement reaction and $S_N 2$ reactions. }
\begin{tabular}{ ||p{3cm}||p{3cm}||p{3cm}|| }

 \hline
 \multicolumn{3}{|c|}{Hyperparameters of SchNetpack models} \\
 \hline
  Parameter & Claisen rearrangment & $S_N 2$ Reaction\\
 \hline
 Representation   & PaiNN & PaiNN   \\
 n\_atom\_basis   &   64 & 128 \\
 n\_interactions & 9   & 9\\
 Radial basis function & Guassian & Guassian\\
 n\_rbf  &   20  & 20\\
 cutoff\_fn & CosineCutoff  & CosineCutoff\\
 cutoff & 10  & 10\\
 loss\_fn & MSE  & MSE\\
 Optimizer & AdamW  & AdamW\\
 Learning rate & 10$^{-4}$ & 10$^{-4}$\\ 
 \hline
\end{tabular}
\end{center}
\end{table}
Different hyperparameters were tested starting from the default values of PaiNN and changing them such that a good compromise between accuracy and computational efficiency could be achieved.
In SchNetPack, interaction layers are followed by a series of atom-wise dense layers \cite{schutt2018schnet} where the following layer is compromised of half the number of neurons of the previous until 1 is reached, i.e. 64, 32, 16, 8, 4, 2, 1. 

\begin{figure}[hbt!]
  \centering
    \centering
    \includegraphics[width=\textwidth]{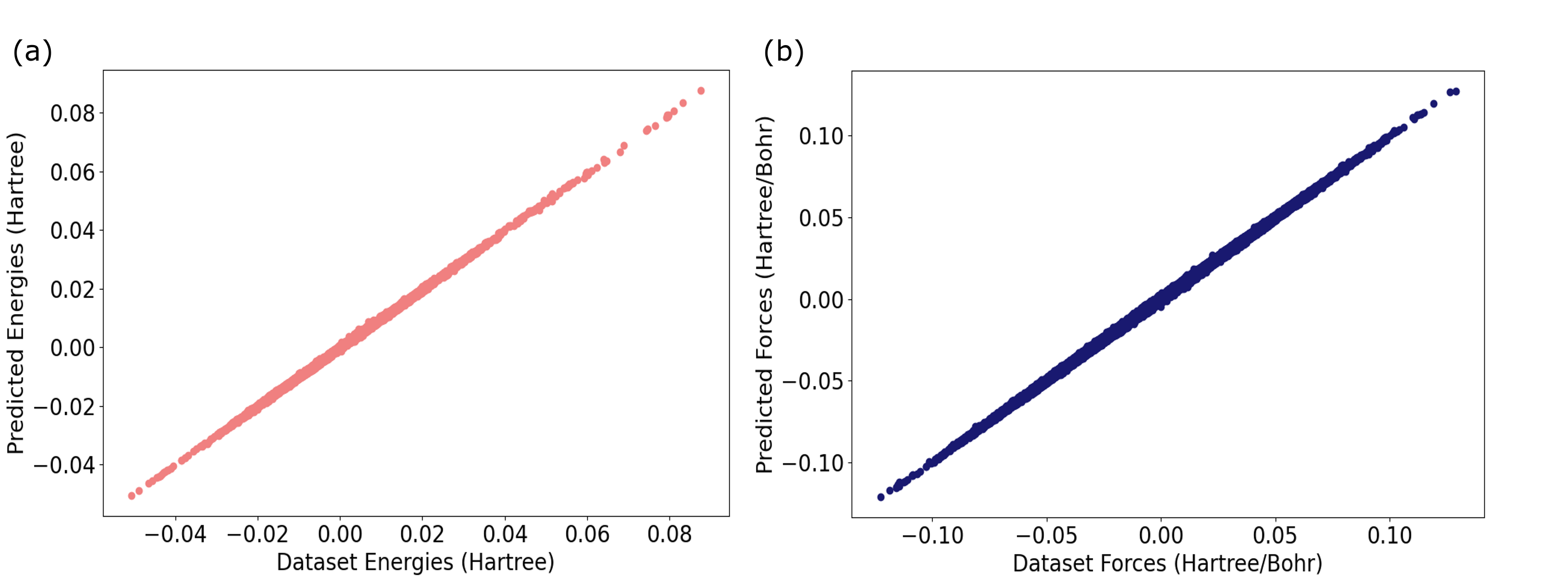}
    \caption{Claisen rearrangement reaction dataset: Scatter plot of a) energies and b) forces showing the test data against the values predicted by the PaiNN model.}
    \label{fig:claisen_painn}
\end{figure}

\begin{figure}[hbt!]
  \centering
    \centering
    \includegraphics[width=\textwidth]{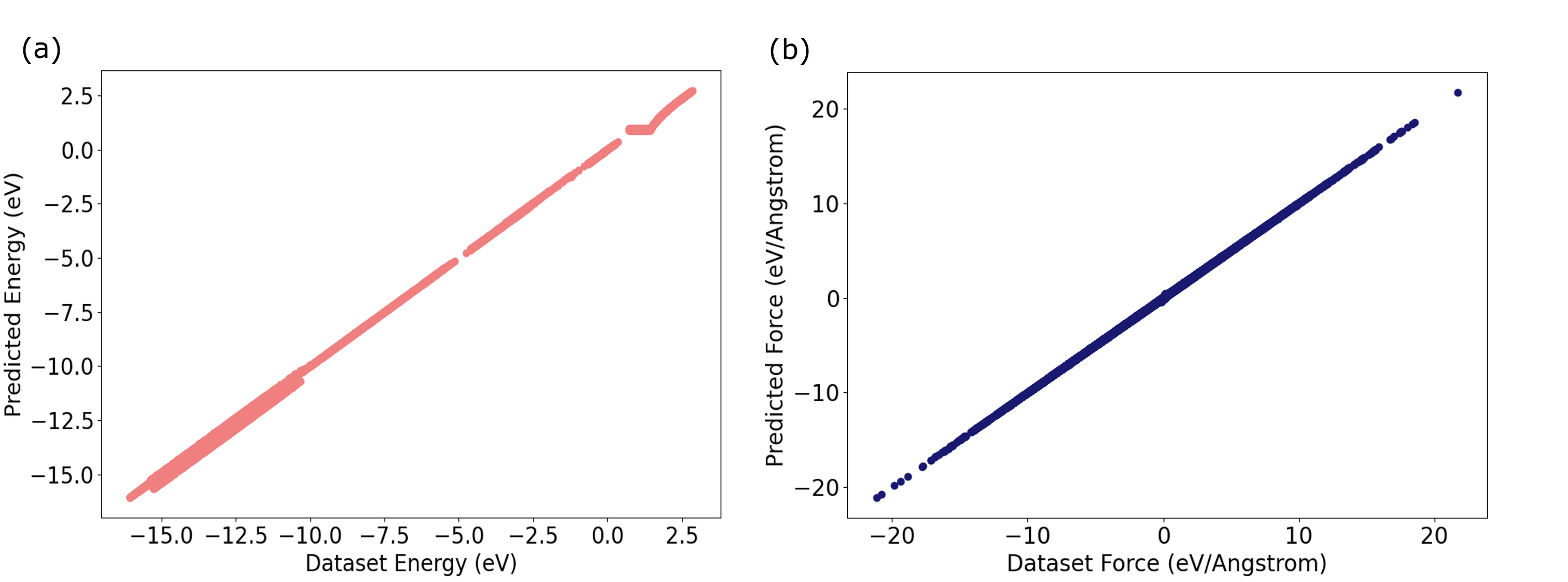}
    \caption{$S_{N}2$ dataset: Scatter plots of a) energies and b) forces showing the test dataset against the values predicted by the PaiNN model.}
    \label{fig:SN2_painn}
\end{figure}

Figure \ref{fig:claisen_painn} and \ref{fig:SN2_painn} show the energies and forces in the test dataset against the energies and forces predicted by the PaiNN \cite{schutt2021equivariant} models respectively.
\subsection{Claisen Rearrangement Datset Error Graph}
The test set consists of 5000 data points. The original data set contains energies being measured in Hartree and forces measured in Hartree/Bohr. The mean absolute errors (MAEs) of energies and forces are 0.0003438 Ha and 0.000316 Ha/Bohr. 
\subsection{$S_N 2$ Dataset Error Graph}
The test set consists of 52709 datapoints with the energies being measured in eV and the forces measured in eV/\AA. The MAEs of energies and forces are 0.03756 eV and 0.004517 eV/\AA.
\subsection{Model Architecture and Training Details}
The model utilizes the pretrained PaiNN representation for each dataset, which were introduced in the previous section, as an initial representation for the model. These representation vectors remain constant throughout the training episodes of the model.

\begin{figure}[hbt!]
\includegraphics[width=\textwidth]{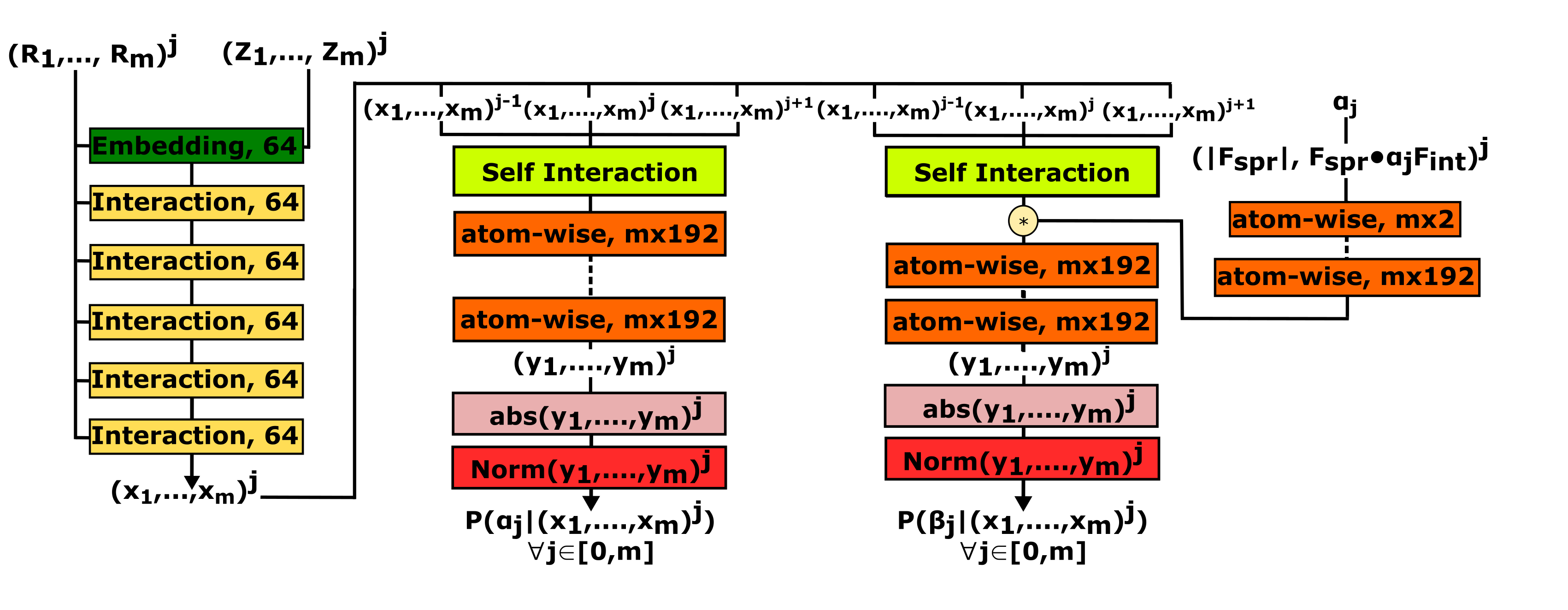}
\caption{\textbf{Actor architecture.} The PaiNN representation forms the initial feature vectors for the actor. The positions of each atom in the current image are passed through a self interaction layer and a dense layer to produce the $\alpha_i$ distribution. A similar process is used to construct the associated $\beta_i$ distribution. The two distributions are then sampled from to produce the change in positions at the current time step.}
\label{fig:actor}
\end{figure}

We show the actor architecture again in Figure \ref{fig:actor} for better understanding of the following text. The self-interaction layers of the model are composed of a multi-attention head \cite{vaswani2017attention}, with one head specifically designed to combine neighboring image representation vectors. The atom-wise dense layers in the model follow from the self-interaction layers, maintaining the same construction methodology discussed in the previous section; the following dense layer being composed of half the number of neurons in the previous layer. The output values were then sampled from the generated distributions. All energies and forces throughout were predicted using the PaiNN model from the previous section. 

To streamline the performance, the spring force vector is used. The spring force vector is precalculated as an input into the model. In addition, the internal force vector is derived using the pretrained PaiNN model, serving as an additional input to the model. This incorporation of both force vectors reduces the computational time needed for training.
For the Claisen rearrangement reaction and the $S_{N}2$ reactions, the model utilizes specific hyperparameters, which are provided in Table 2.
\begin{table}[hbt!]
\begin{center}
\caption{Hyperparameters used for the model when training on the Claisen rearrangement reaction and $S_N 2$ reactions.}
\begin{tabular}{ ||p{3cm}||p{3cm}||p{3cm}|| }

 \hline
 \multicolumn{3}{|c|}{Hyperparameters of the model} \\
 \hline
  Parameter & Claisen rearrangement reaction & $S_N 2$ reactions\\
 \hline
 Batch size   &   10 & 20 \\
 Grid size & $10 \times 10$  & $10 \times 10$  \\
 $\alpha, \beta$ range & 100 & 3 \\ 
 Spring constant (\textit{k}) & 0.1 & 0.1 \\
 Learning rate & 10$^{-4}$ & 10$^{-4}$ \\
 Preoptimise length & 50 & 0 \\
 Episode length & 10 & 10 \\
 Optimizer & Adam & Adam \\
 \hline
\end{tabular}
\end{center}
\end{table}
During the training period two pathways found by the model are stored. These are the pathway corresponding to the minimum activation energy and minimum $F_{max}$ found through all episodes. Due to the stochastic sampling procedure it is worth to mention that results are likely not reproducible exactly, but within a certain statistical accuracy. For better reproducibility, however, the associated trained PaiNN models are provided in addition to this document and code. 

\section{Details of Implemented Nudged Elastic Band (NEB) Method}
The reaction pathway is a crucial concept in understanding chemical reactions. It provides insights into the sequence of structural changes occurring between the initial and final states of a reaction. In order to construct a reaction pathway, a method called the nudged elastic band (NEB) \cite{jonsson1998nudged} is commonly used.

The reaction pathway is formed starting from the start and end structures and using a series of $N+1$ images connected by virtual spring forces forming an elastic band. The elastic band with $N+1$ images can be written as $ \{ \textbf{R}_0, \textbf{R}_1 , ..., \textbf{R}_{n-1}, \textbf{R}_n \}$ where $\textbf{R}_0$ and $\textbf{R}_n$ are the initial and final structures respectively. The initial and final images correspond to specific molecular configurations, which correspond to local energy minima. The positions of the endpoints are thus held fixed to maintain a constant energy throughout the process.

To connect these images and simulate the structural changes along the pathway, virtual spring forces are introduced. The key component of these spring forces is the tangent vector, denoted as $\tau_i$, which describes the direction of the pathway at each image. Initially, a simple estimate of the tangent vector can be obtained by calculating the normalized line segments between neighboring images and summing them:

\begin{equation}
    \tau_i = \frac{\textbf{R}_i - \textbf{R}_{i-1}}{|\textbf{R}_i - \textbf{R}_{i-1}|} + \frac{\textbf{R}_{i+1} - \textbf{R}_{i}}{|\textbf{R}_{i+1} - \textbf{R}_{i}|}
\end{equation}

However, this basic approach can lead to the formation of kinks or irregularities in the pathway. More information about this can be found in reference \citenum{jonsson1998nudged}. To address this issue, a more sophisticated method is proposed. Let $i_{max}$ be the index of the image with the maximum potential energy; this corresponds to the transition state.

\begin{equation}
\begin{gathered}
\tau_i^{-} = \frac{\textbf{R}_i - \textbf{R}_{i-1}}{|\textbf{R}_i -\textbf{R}_{i-1}|} \:\:\:\:\: i < i_{max} \\
\tau_i^{+} = \frac{\textbf{R}_{i+1} - \textbf{R}_{i}}{|\textbf{R}_{i+1} - \textbf{R}_{i}|} \:\:\:\:\: i > i_{max}  \\
\tau_{i_{max}} = \frac{\textbf{R}_i - \textbf{R}_{i-1}}{|\textbf{R}_i -\textbf{R}_{i-1}|} + \frac{\textbf{R}_{i+1} - \textbf{R}_{i}}{|\textbf{R}_{i+1} - \textbf{R}_{i}|} \:\:\:\:\: i = i_{max}
\end{gathered}
\end{equation}
With the tangent vectors established, the respective spring forces between neighboring images can be calculated. Additionally these forces are parameterized by a spring constant, denoted as k, which quantifies the strength of the interaction between neighboring images. The spring force denoted as $\textbf{F}^{i}_{spr} \parallel$, can be expressed as:
\begin{equation}
    \textbf{F}^{i}_{spr} \parallel = k(| \textbf{R}_{i+1} - \textbf{R}_{i}| - |\textbf{R}_{i} - \textbf{R}_{i-1}|)\tau_i
\end{equation}
Additionally, the perpendicular force, denoted as $\textbf{F}_{int}\perp$, which arises from internal molecular forces is considered. This term is defined as the internal force vector that acts orthogonal to the spring force vector. The combined force acting on each image, denoted as $\textbf{F}^{i}$, is then defined as the sum of the spring force and the internal molecular force acting perpendicular to the pathway:
\begin{equation}
\begin{gathered}
\textbf{F}^{i} = \textbf{F}^{i}_{spr} \parallel + \textbf{F}_{int}\perp \\ \text{with  }
\textbf{F}_{int}\perp = \nabla E(R_i) \perp = \nabla E(R_i) - \nabla E(R_i) \parallel .
\end{gathered}
\end{equation}
After each episode, molecular positions are then optimised using an optimisation algorithm. In the NEB experiments performed optimisation was done using BFGS \cite{yuan1991modified}. By applying the BFGS algorithm to the force vector, one can iteratively adjust the positions of the images along the reaction coordinate, seeking the minimum energy pathway. The algorithm uses the current force vector and the approximated Hessian matrix to determine the search direction and step length for each iteration. The positions of the images are updated accordingly, aiming to converge towards a pathway with lower reaction barrier. More information on this can be found in \citenum{yuan1991modified}. The model constructs a new forces vector with the predicted $\alpha_i$ and $\beta_i$ values;
\begin{equation}
    F^{i} = \alpha_i F_{int}\perp + \beta_i F^{i}_{spr} \parallel 
\end{equation}
one step is then taken with the BFGS algorithm to generate the new positions given the relevant force vectors. The BFGS algorithm and script to calculate $F^i$ values given the $\alpha_i$ and $\beta_i$ used the Atomic Simulation Environment (ASE). \cite{larsen2017atomic}

\section{Quantum chemical reference calculations}
All quantum chemical calculations were conducted using the ORCA 5.0.3 \cite{neese2012orca} program suite. For SCF cycles the convergence levels were set to tight SCF and the integration grid was set to defgrid2. In the dataset for the Claisen rearrangement reaction all calculations were conducted using the TZVP basis set and the PBE0 \cite{adamo1999toward} functional. For the $S_N 2$ dataset all calculations were performed with the TZVP basis set, the DSD-BLYP functional \cite{kozuch2010dsd} and the D3BJ \cite{goerigk2017comprehensive} dispersion correction. The details for methods used in constructing the respective datasets for the Claisen rearrangement reaction and $S_N 2$ reactions can be found at \cite{gastegger2021machine} and \cite{unke2019physnet} respectively.

The minimum energy pathway for the Claisen rearrangement reaction was calculated using the NEB implementation in ORCA. Initially the NEB was run with the PBE functional \cite{ernzerhof1999assessment} and SVP basis set. This pathway was used as a starting point for the following NEB. This NEB was done using the TZVP basis set with the PBE0 functional and D4 dispersion correction \cite{caldeweyher2019generally}. The NEB calculations were run using the climbing image implementation \cite{henkelman2000climbing}. 

To calculate the root mean squared deviation (RMSD) between the generated transition states and the reference transition states we did a transition state optimisation with DFT. For the $S_N 2$ reactions, predicted transition states from the model were taken as the initial starting points for transition state optimisation. An eigenvalue following algorithm was applied using OptTS option in ORCA. The TZVP basis set was used along with the RI-C auxiliary basis set \cite{aquilante2007unbiased}. Again the DSD-BLYP functional and the D3BJ dispersion correction were used.

\subsection{Frequency Calculations}

The frequency calculations of generated structures were performed using the DSD-BLYP functional, D3BJ dispersion correction, TZVP basis set and RI-C auxiliary basis set. The frequency calculations were computed numerically.
The associated frequencies for each of the $S_N 2$ reactions of the following form can be seen in the table below. $$X^{\text{–}} + H_3C\text{–}Y \rightarrow X\text{–}CH_3 + Y^\text{–} \:\:\:\: X, Y \in \{F, Cl, Br, I\}$$ 

\begin{table} [hbt!]
\caption{Normal modes of the found transition state for the $S_N 2$ reactions.}
\begin{tabular}{ ||p{3cm}||p{3cm}||p{3cm}||p{3cm}|| }
 \hline
 \multicolumn{4}{|c|}{$S_{N}2$ Reactions Frequencies} \\
 \hline
 Degrees of Freedom & $X=Cl$, $Y=Br$ [cm$^{-1}$] &$X=Cl$, $Y=I$ [cm$^{-1}$]&$X=Br$, $Y=I$ [cm$^{-1}$]\\
 \hline
 0   & 0.0    & 0.0 &   0.0\\
 1 &   0.0  & 0.0   &0.0\\
 2 & 0.0 &  0.0 & 0.0\\
 3    &0.0 & 0.0&  0.0\\
 4 &   0.0  & 0.0&0.0\\
 5 & 0.0  & 0.0   &0.0\\
 6 & -795.92  & -718.09 & -604.47\\
 7 & 142.54  & 132.18 & 104.66\\
 8 & 149.33  & 157.69 & 134.27\\
 9 & 150.75  & 160.16 & 137.94\\
 10 & 791.66  & 801.02 & 801.50\\
 11 & 792.68  & 802.21 & 803.23\\
 12 & 895.63  & 907.19 & 885.53\\
 13 & 1387.45  & 1392.63 & 1390.87\\
 14 & 1387.92  & 1394.92 & 1391.26\\
 15 & 3352.51  & 3336.25 & 3338.38\\
 16 & 3575.85  & 3553.72 & 3560.16\\
 17 & 3577.45  & 3556.60 & 3560.77\\

 \hline
\end{tabular}
\end{table}
The frequency calculation for the transition state of the Claisen rearrangement reaction generated by the model was performed with the TZVP basis set and the PBE0 functional. As with the $S_N 2$ reaction all frequencies were computed numerically. The molecules has 63 normal modes hence results will be provided in an additional document. The structure has 6 negative frequencies, which reveals that the structure is not exactly the transition state. However, most frequencies are very small in their magnitude, i.e., only two are smaller than -300, which indicates that the structure is close to the transition state. In fact, many transition states found with NEB and quantum chemistry reference methods often comprise of more than one negative frequency. The RMSD of the calculated transition state and the transition state obtained with NEB using DFT is about 0.79 \AA which is reasonable given the many degrees of freedom of the molecule.